\documentclass[twocolumn, notitlepage, secnumarabic,amssymb, aps, prl, figure, superscriptaddress, floatfix]{revtex4-1}

\usepackage{graphicx} 
\usepackage{amsmath}
\usepackage{subfigure}
\usepackage{braket}
\usepackage{hyperref}
\usepackage{bm}
\hypersetup{
    colorlinks=true,
    linkcolor=blue,
    filecolor=red,      
    urlcolor=blue,
    citecolor=magenta
}
\usepackage[version=4]{mhchem}
\usepackage{ragged2e}
\usepackage{float}

\usepackage[usenames, dvipsnames]{color}
\definecolor{mypink}{RGB}{219, 48, 122}
\begin{document}

\title{A Bayesian Committee Machine Potential for Organic Nitrogen Compounds}

\author{Hyun Gyu Park}
\affiliation{Department of Energy Science, Sungkyunkwan University, Seobu-ro 2066, Suwon, 16419, Korea}

\author{Soohaeng Yoo Willow}
\affiliation{Department of Energy Science, Sungkyunkwan University, Seobu-ro 2066, Suwon, 16419, Korea}

\author{D. ChangMo Yang}
\email{dcyang@skku.edu}
\affiliation{Department of Energy Science, Sungkyunkwan University, Seobu-ro 2066, Suwon, 16419, Korea}

\author{Chang Woo Myung}
\email{cwmyung@skku.edu}
\affiliation{Department of Energy Science, Sungkyunkwan University, Seobu-ro 2066, Suwon, 16419, Korea}

\date{\today}

\begin{abstract}
Large-scale computer simulations of chemical atoms are used in a wide range of applications, including batteries, drugs, and more. However, there is a problem with efficiency as it takes a long time due to the large amount of calculation. To solve these problems, machine learning interatomic potential (ML-IAP) technology is attracting attention as an alternative. ML-IAP not only has high accuracy by faithfully expressing the density functional theory (DFT), but also has the advantage of low computational cost. However, there is a problem that the potential energy changes significantly depending on the environment of each atom, and expansion to a wide range of compounds within a single model is still difficult to build in the case of a kernel-based model. To solve this problem, we would like to develop a universal ML-IAP using this active Bayesian Committee Machine (BCM) potential methodology for carbon-nitrogen-hydrogen (CNH) with various compositions. ML models are trained and generated through first-principles calculations and molecular dynamics simulations for molecules with only CNH. Using long amine structures to test an ML model trained only with short chains, the results show excellent consistency with DFT calculations. Consequently, machine learning-based models for organic molecules not only demonstrate the ability to accurately describe various physical properties but also hold promise for investigating a broad spectrum of diverse materials systems. 
\end{abstract}
\maketitle
\section{Introduction}
Rapid advances in CPU and GPU parallelism as well as high-throughput cluster network architectures have enabled researchers and corporations worldwide to employ numerous algorithmic innovations, resulting in a variety of applications in biology, physics, climate science, and chemistry at multiple scales~\cite{alder1959studies, mccammon1977dynamics, le2023symmetry}.
Traditional methods such as Optimized Potentials for Liquid Simulations (OPLS)~\cite{jorgensen1984optimized}, Chemistry at Harvard Macromolecular Mechanics (CHARMM)~\cite{brooks2009charmm}, Assisted Model Building with Energy Refinement (AMBER)~\cite{cornell1995second}, embedded-atom method (EAM)~\cite{daw1983semiempirical}, reactive bond-order (\textsc{ReaxFF})~\cite{van2001reaxff}, charge-optimized many-body (COMB)~\cite{liang2012variable}, and other widely used classical force-fields or interatomic potentials enable large-scale simulations that are not feasible with \textit{ab initio} calculations.
They also have the advantage of reproducing some experimental data~\cite{hajibabaei2021machine}. However, classical force-fields (FF) frequently demonstrate limitations in representing helicity and noncovalent intermolecular interactions~\cite{hajibabaei2021machine, tzanov2014accurately}. Additionally, they often struggle to accurately describe correct structures and all associated properties~\cite{kondratyuk2016self, kondratyuk2019comparing}. Consequently, their accuracy is constrained by these shortcomings, particularly in depicting various properties with correct structures simultaneously~\cite{ha2022sparse}. Moreover, fitting reactive force-fields necessitates prior knowledge of the reaction networks to be simulated, significantly constraining predictive capability and potentially introducing human bias regarding the reactions that proceed~\cite{zhang2023exploring}. 
An alternative approach, utilizing \textit{ab initio} data-driven ML-IAP methods, has been developed to address the issues and limitations of the FF method. This enables the accurate representation of the potential energy surface (PES)~\cite{unke2021machine, deringer2021gaussian}. Over the last two decades, the application of ML-IAP in materials simulations has been proven highly successful, aiming to address the gap in speed, accuracy, and generality~\cite{zhang2023exploring}. These include various machine learning techniques such as Neural Networks (NNs)~\cite{behler2007generalized}, Artificial Neural Network Methods (ANN)~\cite{pun2019physically}, Graph Neural Network Potentials (GNN)~\cite{batzner20223}, Gaussian Approximation Potentials (GAP)~\cite{bartok2010gaussian}, Gradient Domain Machine Learning (GDML)~\cite{chmiela2017machine}, and Sparse Gaussian Process Regression (SGPR)~\cite{hajibabaei2021sparse}. These methodologies have gained considerable traction not only in the field of physical chemistry but also in various other applied disciplines~\cite{hajibabaei2021sparse, hong2021first}. While NNs are particularly effective at handling broad data sets, they have a huge number of optimizable parameters and require big data to avoid overfitting~\cite{hajibabaei2021universal}. In contrast, Kernel-based methods show excellent performance in training and adapting on the fly using smaller datasets. However, the computational cost for training scales by \(O(n^3)\) depending on the data size $n$~\cite{hajibabaei2021sparse, hajibabaei2021universal}. Consequently, to improve efficiency, the SGPR technique has been developed, utilizing subsets of data~\cite{hajibabaei2021sparse}. However, SGPR continues to face difficulties with a large number of inducing points $m$, exhibiting computational demands of  \(O(nm^2)\)~\cite{willow2024bayesian,willow2024sparse}.
 To overcome these challenges, researchers are exploring aggregation models such as Bayesian Committee Machines (BCMs)~\cite{willow2024bayesian,willow2024sparse,hinton2002training, tresp2000bayesian}. The BCM model combines predictions from submodels trained on different sections using SGPR with computational scaling of \(O(nm^2)\). Furthermore, it is anticipated that the scalability of ML-IAP will be enhanced through the extension of existing Gaussian Process (GP) learning methodologies, employing the Bayesian approach.
 In this study, we propose a highly reliable ML-IAP for organic compounds containing carbon, nitrogen, and hydrogen, utilizing the SGPR algorithm and the BCM model. The training set consists of compounds containing only carbon, nitrogen, and hydrogen, such as amines, alkaloids, azoles, cyanides, hydrazines, imidazoles, nitriles and pyridines. The model was tested using amine long chains (\ce{C21NH45}), 
 demonstrating high accuracy despite being trained on short-chain amines. The results of physical property calculations using ML-IAP were similar to those obtained from DFT, confirming the excellence of the SGPR algorithm and BCM model. Our approach represents a significant advancement in the development of general-purpose ML-IAPs for organic compounds.

\section{Theory\label{sec:theory}}

\subsection{Sparse Gaussian Process Regression potential}

ML-IAPs are commonly defined by incorporating all atoms within a defined cutoff radius. In a configuration $x$ consisting of $N$ atoms, a series of descriptors is generated, represented as \( \text{descriptor } x = \{\rho_i\}_{i=1}^N \), with each \( \rho_i \) dependent on the local chemical environment (LCE) of atom $i$ within the cutoff radius. The ML-IAP energy becomes additive across local chemical environment (LCE) \( \rho_i \) as
\begin{eqnarray}
E(\bm{x}) & = & \sum_{i=1}^N \epsilon(\rho_i).
\end{eqnarray}
where \( \epsilon \) represents a fictional energy arising from the interactions between atom $i$ and its neighboring atoms.
In kernel-based regression methods, \(\epsilon \) is expressed as
\begin{eqnarray}
\epsilon(\bm{\rho}) & = & \sum_{j=1}^{m} K(\rho_i, \chi_j) \omega_j,
\end{eqnarray}
where the matrices are represented with bold fonts, \( z = \{\chi_j\}_{j=1}^m \) denotes the set of reference/inducing descriptors, \( w \) represents the vector of weights for the inducing descriptors, and \(K \) signifies a similarity/covariance kernel. The weights are determined to ensure the accurate reproduction of potential energy and forces for a given set of \textit{ab initio} data \( X = \{x_k\}_{k=1}^n \), and their calculation is contingent upon the regression algorithm employed. In SGPR~\cite{hajibabaei2021sparse},

\begin{eqnarray}\label{eq:w}
\bm{w} & = & (\sigma^2\bm{k}_{mm}+\bm{k}^T_{nm}\bm{k}_{nm})^{-1} \bm{k}^T_{nm}\bm{Y}
\end{eqnarray}

\(\bm{k}_{mm}\) and \(\bm{k}_{nm}\) denote the interinducing and data-inducing covariance matrices, respectively. \(\bm{Y}\) signifies the data potential energies (and forces), while \(\sigma\) represents the noise hyperparameter. The noise scale \(\sigma\), along with any hyperparameters within the kernel \(\bm{K}\), are fine-tuned to maximize the likelihood of the energy data. In defining the similarity kernel, we employ a modified version of the smooth overlap of atomic positions (SOAP)~\cite{hajibabaei2021sparse, bartok2013representing}. When utilizing existing data, the inducing descriptors are drawn from the dataset. However, in on-the-fly learning, both the data and inducing descriptors are sampled simultaneously during molecular dynamics (MD). Bayesian methods utilize the predictive variance as a threshold for sampling new data or inducing descriptors. Within SGPR, the predictive variance for \(\epsilon(\rho\)) is determined as

\begin{align} 
\ensuremath{\nu}(\rho) & = \bm{k}_{\rho\rho} - \bm{k}_{\rho m} \bm{k}_{mm}^{-1} \bm{k}_{\rho m}^T \nonumber \\
& \quad + \sigma^2 \bm{k}_{\rho m} (\sigma^2 \bm{k}_{mm} + \bm{k}_{nm}^T)^{-1} \bm{k}_{\rho m}^T \label{eq:nu}
\end{align}


To compute \( w \) at Eq. (\ref{eq:w}), we can transform this equation into a linear system to bypass the matrix inversion \((\sigma^2\bm{k}_{mm} + \bm{k}^T_{nm}\bm{k}_{nm})^{-1}\)~\cite{hajibabaei2021sparse}. However, in Eq. (\ref{eq:nu}), this matrix inversion is explicitly required, a process that can be both numerically unstable and computationally expensive, particularly for on-the-fly learning~\cite{williams2002observations}. Hence, in certain approximations, the third term in this equation is disregarded, potentially leading to slightly suboptimal sampling outcomes. 

\subsection{Bayesian Committee Machine potential}
\begin{figure*} [htp]
\includegraphics[width=6.in]{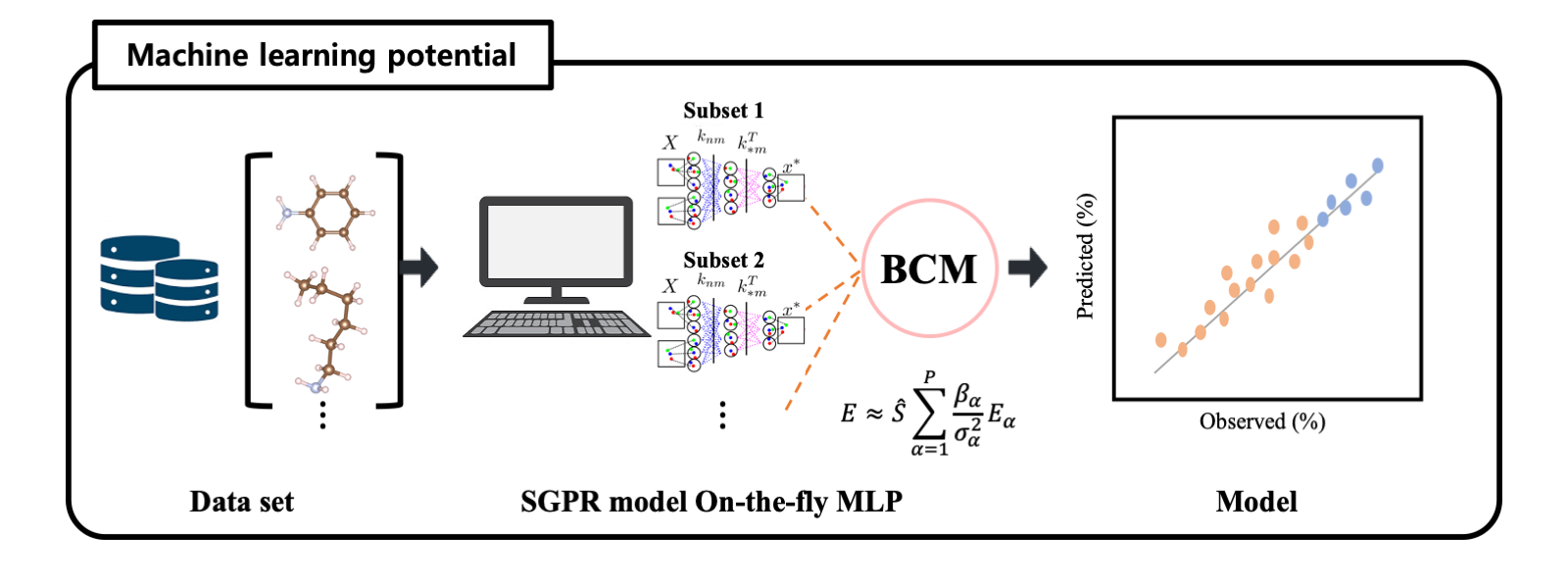}
\caption{The schematic represents the Bayesian committee machine potential. Following the division of the entire dataset into distinct subsets based on classification, training of SGPR-based MLPs occurs within each subset. Through this process, estimation of the universal MLP for the entire dataset is facilitated via the BCM.}
\label{fig:schematic}
\end{figure*}

 For the purpose of enhancing memory efficiency and improving the parallelization of the SGPR algorithm, we adopted the BCM~[30] as an innovative method to amalgamate SGPR-MLPs, each of which was trained independently on distinct subsets, as depicted in Fig. \ref{fig:schematic}. By leveraging the BCM algorithm, we approximate the predicted potential energy as follows:
\begin{eqnarray}
 E & \approx & \hat{S}\frac{\beta_\alpha}{\sigma^2_{\alpha}}E_\alpha
\end{eqnarray}

The calculation of the potential energy \( E_\alpha \) involves the utilization of the \(\alpha\)-th subset (with \( R_{n_{\in \alpha}} \) and \( z_{\alpha} = \{\chi_j\}_{j\in\alpha} \)), along with the weight vector \( w_\alpha \) corresponding to the \( \alpha \)-th sub SGPR model. This energy is described as:
\begin{eqnarray}
 E_{\alpha}(R) \approx \sum_{i, (j \in m_{\alpha})} w_{\alpha j} K(\rho_i, \chi_j).
\end{eqnarray}
We formulate the active BCM potential by selectively sampling the LCE, focusing on instances where the covariance loss \( s(\rho_i) \) exceeds a predefined threshold.
\begin{eqnarray}
 s(\rho_i) = K(\rho_i, \rho_i) - \bm{k}_{\rho_i m} \bm{k}_{mm}^{-1} \bm{k}_{\rho_i m}^T
\end{eqnarray}
The weighting of the \( \alpha \)-th committee prediction relies on the utilization of the maximum covariance loss value \( \sigma^2_\alpha \) for the \( \alpha \)-th subset. 
\begin{eqnarray}
 \sigma^2_{\alpha} = \max \left(1 - \bm{k}_{\rho_i m} \bm{k}_{mm}^{-1} \bm{k}_{\rho_i m}^T\right).
\end{eqnarray}

Moreover, we incorporate a weighting factor \( \beta_\alpha \) inspired by the robust BCM approach, where each committee prediction is scaled by the differential entropy \( \beta_\alpha = - \log(\sigma^2_\alpha) \). The final normalization factor \(\hat{S}\) is thus given:
\begin{eqnarray}
\hat{S} = \left[ \sum\limits_{\alpha}^{p} \frac{\beta_{\alpha}}{\sigma^2_{\alpha}} \right]^{-1}
\end{eqnarray}
The BCM combined with SGPR provides a significant advantage by allowing operation with a reduced (n/p)-dimensional training dataset and an (m/p)-dimensional inducing set, as opposed to the original n-dimensional training dataset and m-dimensional inducing set, where \( n = \sum_{\alpha} n_\alpha \) and \( m = \sum_{\alpha} m_\alpha \). 
As a result, the computational expense in BCM-based SGPR shifts from \( O(nm^2) \) to \( p O(\frac{nm^2}{p^3}) \). Additionally, the BCM offers the advantage of giving more weight to a particular subset that closely resembles a test configuration \( x^* \), thereby exerting a more substantial impact on energy, forces, and stress.

\subsection{Computational Details}
All \textit{ab initio} calculations are conducted using the Vienna \textit{ab initio} simulation package (\textsc{VASP}) version~6.0~\cite{kresse1996efficient}, which employs the projector augmented-wave (PAW)~\cite{blochl1994projector} method within the framework of density functional theory (DFT). The calculations utilize the Perdew-Burke-Ernzerhof (PBE) generalized gradient approximation (GGA) functionals~\cite{perdew1996generalized}.

The kinetic-energy cutoff was set to 500~eV. The convergence criterion for the electronic energy difference was set to \(10^5~\text{eV}\). The Brillouin zone was sampled using the \(\Gamma\)-point. 
Using the atomic simulation environment package (\textsc{ASE})~\cite{larsen2017atomic}, NVT MD simulations were performed using Nos\'{e}--Hoover thermostat and Parrinello-Rahman dynamics to explore different conformations of a given molecule. The MD simulations were conducted for 3-6 ps at 300 K, utilizing a time step of 0.5~fs. To assess the accuracy of ML-IAPs, we employ the coefficient of determination, denoted by \( R^2 \), which is defined as follows:
\begin{equation}
R^2 = 1 - \frac{\sum_{i} (f_{i} - \tilde{f}_{i})^2}{\sum_{i} (f - \bar{f}_{i})^2}
\end{equation}
The Python package \textsc{AutoForce}~\cite{AutoForce}, integrated with the \textsc{ASE}, was employed for on-the-fly generation of SGPR models and learning of the PES.
\begin{figure*} [htp]
\includegraphics[width=6.in]{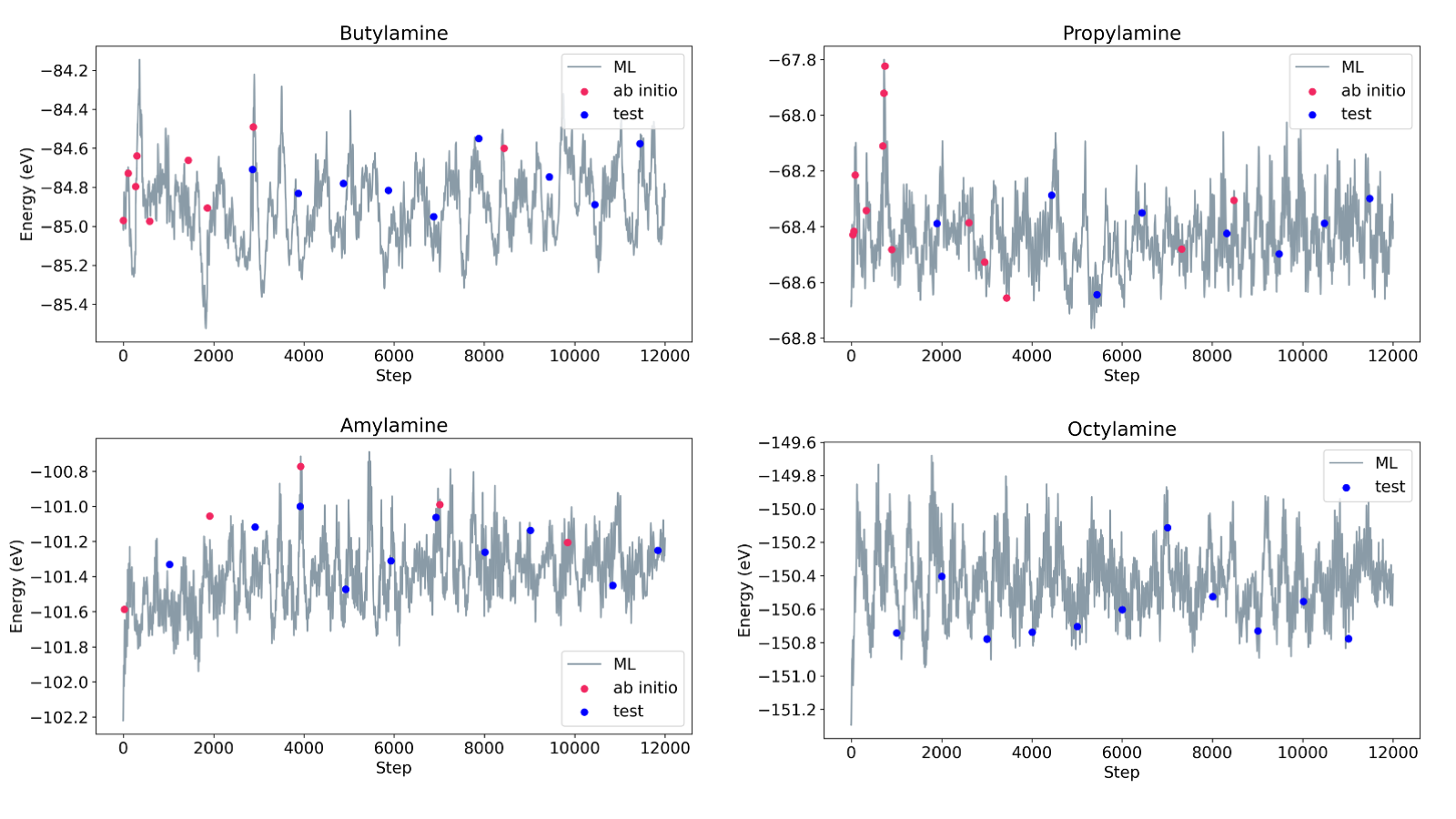}
\caption{Initial steps for on-the-fly MLMD for elementary amine molecules. Red bullets indicate the \textit{ab initio} potential energy of the configurations sampled by the SGPR model. Blue bullets are for single-point \textit{ab initio} tests.}
\label{fig:mlmd}
\end{figure*}

\section{Results}

\begin{figure*} [htp]
\includegraphics[width=6.in]{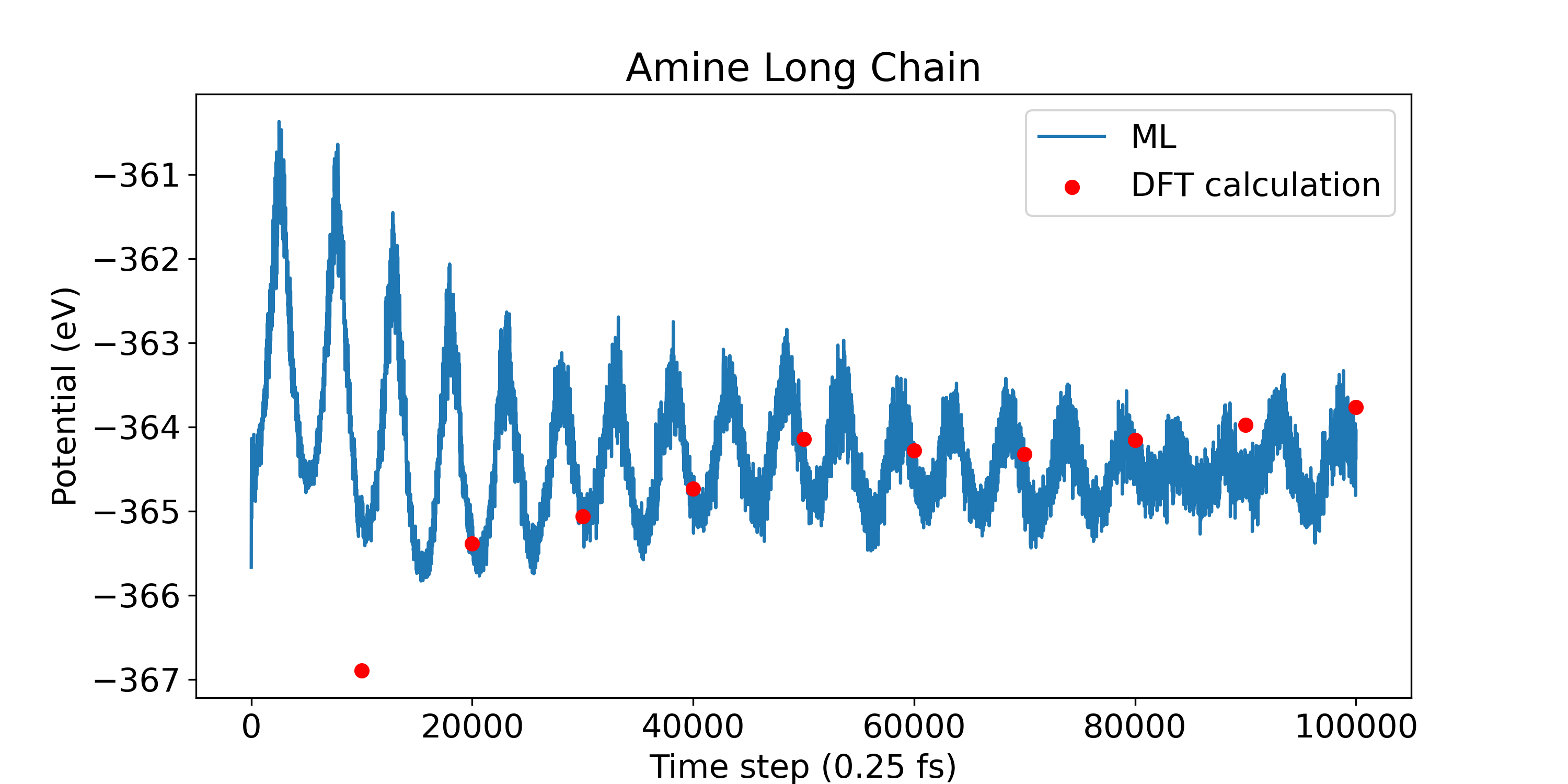}
\caption{Initial steps for on-the-fly MLMD for elementary amine molecules. Red bullets indicate the \textit{ab initio} potential energy of the configurations sampled by the SGPR model. Blue bullets are for single-point \textit{ab initio} tests.}
\label{fig:mlmd_long}
\end{figure*}

The ML-IAP model for each molecular category is generated on the fly through a short MD simulation (Total step: 3 ps, step: 0.25 fs) of that molecule at 300 K, as shown in Fig.~\ref{fig:mlmd}. Fig. \ref{fig:mlmd} shows the training procedure for basic amine molecules. After each MD simulation, the model is updated and used as the basis for the next iteration. Among the many data sets in the amine group, ML calculations were performed starting with Butylamine. The ML potential energy is represented by the gray line, the red circle denotes the outcome of DFT calculation, and the blue circle correspondings to the DFT calculation conducted every 10,000 steps to assess the variance with the SGPR ML calculation. 
In Butylamine in Fig.~\ref{fig:mlmd}, many red circles were found in the early steps. This is because the uncertainty value (calculation setting value) was higher than that during ML training. This is the process of training the model as DFT calculations are run. In the middle step, fewer red circles were confirmed than in the initial step. This means that the model is trained in the initial step and the ML calculation is progressing below the calculation setting value. After calculating butylamine, propylamine was trained. Propylamine can be found in many red circles in the initial steps. This is because propylamine has a different structure from butylamine. However, it was confirmed that it gradually decreased as the calculation progressed. Amylamine was calculated after calculating propylamine. Fewer red circles were visible than the other two compounds because the ML model was trained on it. When calculating Octylamine, the red circle did not appear. This is because ML model training went smoothly. After ML training, 981, 17, 52, 39, 17, 34 and 34 \textit{ab initio} samples have been accumulated for amine, azole, alkaloid, cyanide, hydrazine, imidazole and nitrile, respectively. Table I shows the performance of the ML-IAP model on these train sets, showing metrics such as mean absolute error (MAE).
Similar MD simulations are repeated without immediate training to generate independent test sets. To assess the performance of the developed ML-IAP model, MD simulations (without real-time training) were carried out at 300 K using amine long chains (\ce{C21NH45}). The amine group dataset depicted in Fig.~S1(a) was employed to train the ML-IAP model, while Fig.~S1(b) illustrates the amine long chains used for testing the ML-IAP model. The performance of the amine model is summarized in TABLE S1, which shows a high level of transferability. To ensure the thoroughness of the ML-IAP model training, DFT calculations were conducted at regular intervals of 2.5 ps. The MAE was computed for both DFT and ML calculations at each 2.5 ps interval, revealing consistently low values. This observation strongly suggests that the model training process proceeded smoothly and effectively, validating the robustness and reliability of the ML-IAP model. Fig.~\ref{fig:mlmd_long} depicts the MD test outcomes of the ML-IAP model constructed using (\ce{C21NH45}) data. The blue line represents the ML test results, while the red circles denote the DFT calculation results obtained at 2.5 ps intervals. Initially, disparities between the ML test and DFT calculations were observed. However, as the ML test progressed, the discrepancy in total potential energy gradually diminished. This observation confirms the outstanding performance of the ML-IAP model developed using the amine group dataset.

\begin{table}[h!]
\caption{Testing the Expert Models}
\resizebox{\columnwidth}{!}{%
\begin{tabular}{ccccc}
\hline\hline
Group     & Number of test samples & \begin{tabular}[c]{@{}c@{}}Energy MAE\\ (meV)\end{tabular} & \begin{tabular}[c]{@{}c@{}}Force MAE\\ (eV/Å)\end{tabular} \\
\hline\hline
amine     & 981                      & 0.184                                                            & 0.099                                                              &                            \\
azole     & 17                       & 0.036                                                            & 0.091                                                              &                            \\
alkaloid  & 52                       & 0.242                                                            & 0.120                                                              &                            \\
cyanide   & 39                       & 0.052                                                            & 0.099                                                              &                            \\
hydrazine & 17                       & 0.104                                                            & 0.114                                                              &                            \\
imidazole & 34                       & 0.056                                                            & 0.097                                                              &                            \\
nitrile   & 99                       & 0.129                                                            & 0.099                                                               &                           \\
\hline
\end{tabular}%
}
\end{table}

\section{Discussion}
In this study, ML-IAPs acquired through the SGPR algorithm are utilized to predict the molecular behavior of CNH compounds. Expert models are developed for distinct CNH groups via on-the-fly ML, which are subsequently amalgamated to form universal ML-IAPs. These universal ML-IAPs, encompassing crucial potential energy surfaces, demonstrate exceptional transferability across diverse CNH systems. Collectively, these findings underscore the efficient generation of universal models from individually trained atomic potentials targeting various subgroup configurations. This approach presents a novel opportunity to create universal models for multi-component systems, a capability previously unattainable with other ML methodologies. Moreover, the study emphasizes the effectiveness of the combined model (BCM) generated through kernel-based techniques like SGPR, offering heightened generality and transferability. The utilization of BCM signifies a more robust and reliable framework for broader applications within the realm of ML-IAP.

\section{Acknowledgements}
SYW and CWM acknowledge the support from the National Research Foundation of Korea (NRF) grant funded by the Korea government (MSIT) (No. RS2023-00222245). CWM acknowledges the support from the National Research Foundation of Korea (NRF) grant funded by the Korea government (MSIT) (No. NRF-2022R1C1C1010605). The authors are grateful for: the computational support from the Korea Institute of Science and Technology Information (KISTI) for the \textsc{Nurion} cluster (KSC-2023-CRE-0059,KSC-2023-CRE-0355,KSC-2023-CRE-0454).

\bibliography{ref}

\begin{thebibliography}{36}%
\makeatletter
\providecommand \@ifxundefined [1]{%
 \@ifx{#1\undefined}
}%
\providecommand \@ifnum [1]{%
 \ifnum #1\expandafter \@firstoftwo
 \else \expandafter \@secondoftwo
 \fi
}%
\providecommand \@ifx [1]{%
 \ifx #1\expandafter \@firstoftwo
 \else \expandafter \@secondoftwo
 \fi
}%
\providecommand \natexlab [1]{#1}%
\providecommand \enquote  [1]{``#1''}%
\providecommand \bibnamefont  [1]{#1}%
\providecommand \bibfnamefont [1]{#1}%
\providecommand \citenamefont [1]{#1}%
\providecommand \href@noop [0]{\@secondoftwo}%
\providecommand \href [0]{\begingroup \@sanitize@url \@href}%
\providecommand \@href[1]{\@@startlink{#1}\@@href}%
\providecommand \@@href[1]{\endgroup#1\@@endlink}%
\providecommand \@sanitize@url [0]{\catcode `\\12\catcode `\$12\catcode `\&12\catcode `\#12\catcode `\^12\catcode `\_12\catcode `\%12\relax}%
\providecommand \@@startlink[1]{}%
\providecommand \@@endlink[0]{}%
\providecommand \url  [0]{\begingroup\@sanitize@url \@url }%
\providecommand \@url [1]{\endgroup\@href {#1}{\urlprefix }}%
\providecommand \urlprefix  [0]{URL }%
\providecommand \Eprint [0]{\href }%
\providecommand \doibase [0]{http://dx.doi.org/}%
\providecommand \selectlanguage [0]{\@gobble}%
\providecommand \bibinfo  [0]{\@secondoftwo}%
\providecommand \bibfield  [0]{\@secondoftwo}%
\providecommand \translation [1]{[#1]}%
\providecommand \BibitemOpen [0]{}%
\providecommand \bibitemStop [0]{}%
\providecommand \bibitemNoStop [0]{.\EOS\space}%
\providecommand \EOS [0]{\spacefactor3000\relax}%
\providecommand \BibitemShut  [1]{\csname bibitem#1\endcsname}%
\let\auto@bib@innerbib\@empty
\bibitem [{\citenamefont {Alder}\ and\ \citenamefont {Wainwright}(1959)}]{alder1959studies}%
  \BibitemOpen
  \bibfield  {author} {\bibinfo {author} {\bibfnamefont {B.~J.}\ \bibnamefont {Alder}}\ and\ \bibinfo {author} {\bibfnamefont {T.~E.}\ \bibnamefont {Wainwright}},\ }\href@noop {} {\bibfield  {journal} {\bibinfo  {journal} {J.~Chem.~Phys.}\ }\textbf {\bibinfo {volume} {31}},\ \bibinfo {pages} {459} (\bibinfo {year} {1959})}\BibitemShut {NoStop}%
\bibitem [{\citenamefont {McCammon}\ \emph {et~al.}(1977)\citenamefont {McCammon}, \citenamefont {Gelin},\ and\ \citenamefont {Karplus}}]{mccammon1977dynamics}%
  \BibitemOpen
  \bibfield  {author} {\bibinfo {author} {\bibfnamefont {J.~A.}\ \bibnamefont {McCammon}}, \bibinfo {author} {\bibfnamefont {B.~R.}\ \bibnamefont {Gelin}}, \ and\ \bibinfo {author} {\bibfnamefont {M.}~\bibnamefont {Karplus}},\ }\href@noop {} {\bibfield  {journal} {\bibinfo  {journal} {Nature}\ }\textbf {\bibinfo {volume} {267}},\ \bibinfo {pages} {585} (\bibinfo {year} {1977})}\BibitemShut {NoStop}%
\bibitem [{\citenamefont {Le}\ \emph {et~al.}(2023)\citenamefont {Le}, \citenamefont {Kiss}, \citenamefont {Schuhmacher}, \citenamefont {Tavernelli},\ and\ \citenamefont {Tacchino}}]{le2023symmetry}%
  \BibitemOpen
  \bibfield  {author} {\bibinfo {author} {\bibfnamefont {I.~N.~M.}\ \bibnamefont {Le}}, \bibinfo {author} {\bibfnamefont {O.}~\bibnamefont {Kiss}}, \bibinfo {author} {\bibfnamefont {J.}~\bibnamefont {Schuhmacher}}, \bibinfo {author} {\bibfnamefont {I.}~\bibnamefont {Tavernelli}}, \ and\ \bibinfo {author} {\bibfnamefont {F.}~\bibnamefont {Tacchino}},\ }\href@noop {} {\bibfield  {journal} {\bibinfo  {journal} {arXiv preprint arXiv:2311.11362}\ } (\bibinfo {year} {2023})}\BibitemShut {NoStop}%
\bibitem [{\citenamefont {Jorgensen}\ \emph {et~al.}(1984)\citenamefont {Jorgensen}, \citenamefont {Madura},\ and\ \citenamefont {Swenson}}]{jorgensen1984optimized}%
  \BibitemOpen
  \bibfield  {author} {\bibinfo {author} {\bibfnamefont {W.~L.}\ \bibnamefont {Jorgensen}}, \bibinfo {author} {\bibfnamefont {J.~D.}\ \bibnamefont {Madura}}, \ and\ \bibinfo {author} {\bibfnamefont {C.~J.}\ \bibnamefont {Swenson}},\ }\href@noop {} {\bibfield  {journal} {\bibinfo  {journal} {J.~Am.~Chem.~Soc.}\ }\textbf {\bibinfo {volume} {106}},\ \bibinfo {pages} {6638} (\bibinfo {year} {1984})}\BibitemShut {NoStop}%
\bibitem [{\citenamefont {Brooks}\ \emph {et~al.}(2009)\citenamefont {Brooks}, \citenamefont {Brooks~III}, \citenamefont {Mackerell~Jr}, \citenamefont {Nilsson}, \citenamefont {Petrella}, \citenamefont {Roux}, \citenamefont {Won}, \citenamefont {Archontis}, \citenamefont {Bartels}, \citenamefont {Boresch} \emph {et~al.}}]{brooks2009charmm}%
  \BibitemOpen
  \bibfield  {author} {\bibinfo {author} {\bibfnamefont {B.~R.}\ \bibnamefont {Brooks}}, \bibinfo {author} {\bibfnamefont {C.~L.}\ \bibnamefont {Brooks~III}}, \bibinfo {author} {\bibfnamefont {A.~D.}\ \bibnamefont {Mackerell~Jr}}, \bibinfo {author} {\bibfnamefont {L.}~\bibnamefont {Nilsson}}, \bibinfo {author} {\bibfnamefont {R.~J.}\ \bibnamefont {Petrella}}, \bibinfo {author} {\bibfnamefont {B.}~\bibnamefont {Roux}}, \bibinfo {author} {\bibfnamefont {Y.}~\bibnamefont {Won}}, \bibinfo {author} {\bibfnamefont {G.}~\bibnamefont {Archontis}}, \bibinfo {author} {\bibfnamefont {C.}~\bibnamefont {Bartels}}, \bibinfo {author} {\bibfnamefont {S.}~\bibnamefont {Boresch}},  \emph {et~al.},\ }\href@noop {} {\bibfield  {journal} {\bibinfo  {journal} {J.~Comput.~Chem.}\ }\textbf {\bibinfo {volume} {30}},\ \bibinfo {pages} {1545} (\bibinfo {year} {2009})}\BibitemShut {NoStop}%
\bibitem [{\citenamefont {Cornell}\ \emph {et~al.}(1995)\citenamefont {Cornell}, \citenamefont {Cieplak}, \citenamefont {Bayly}, \citenamefont {Gould}, \citenamefont {Merz}, \citenamefont {Ferguson}, \citenamefont {Spellmeyer}, \citenamefont {Fox}, \citenamefont {Caldwell},\ and\ \citenamefont {Kollman}}]{cornell1995second}%
  \BibitemOpen
  \bibfield  {author} {\bibinfo {author} {\bibfnamefont {W.~D.}\ \bibnamefont {Cornell}}, \bibinfo {author} {\bibfnamefont {P.}~\bibnamefont {Cieplak}}, \bibinfo {author} {\bibfnamefont {C.~I.}\ \bibnamefont {Bayly}}, \bibinfo {author} {\bibfnamefont {I.~R.}\ \bibnamefont {Gould}}, \bibinfo {author} {\bibfnamefont {K.~M.}\ \bibnamefont {Merz}}, \bibinfo {author} {\bibfnamefont {D.~M.}\ \bibnamefont {Ferguson}}, \bibinfo {author} {\bibfnamefont {D.~C.}\ \bibnamefont {Spellmeyer}}, \bibinfo {author} {\bibfnamefont {T.}~\bibnamefont {Fox}}, \bibinfo {author} {\bibfnamefont {J.~W.}\ \bibnamefont {Caldwell}}, \ and\ \bibinfo {author} {\bibfnamefont {P.~A.}\ \bibnamefont {Kollman}},\ }\href@noop {} {\bibfield  {journal} {\bibinfo  {journal} {J.~Am.~Chem.~Soc.}\ }\textbf {\bibinfo {volume} {117}},\ \bibinfo {pages} {5179} (\bibinfo {year} {1995})}\BibitemShut {NoStop}%
\bibitem [{\citenamefont {Daw}\ and\ \citenamefont {Baskes}(1983)}]{daw1983semiempirical}%
  \BibitemOpen
  \bibfield  {author} {\bibinfo {author} {\bibfnamefont {M.~S.}\ \bibnamefont {Daw}}\ and\ \bibinfo {author} {\bibfnamefont {M.~I.}\ \bibnamefont {Baskes}},\ }\href@noop {} {\bibfield  {journal} {\bibinfo  {journal} {Phys. Rev. Lett.}\ }\textbf {\bibinfo {volume} {50}},\ \bibinfo {pages} {1285} (\bibinfo {year} {1983})}\BibitemShut {NoStop}%
\bibitem [{\citenamefont {Van~Duin}\ \emph {et~al.}(2001)\citenamefont {Van~Duin}, \citenamefont {Dasgupta}, \citenamefont {Lorant},\ and\ \citenamefont {Goddard}}]{van2001reaxff}%
  \BibitemOpen
  \bibfield  {author} {\bibinfo {author} {\bibfnamefont {A.~C.}\ \bibnamefont {Van~Duin}}, \bibinfo {author} {\bibfnamefont {S.}~\bibnamefont {Dasgupta}}, \bibinfo {author} {\bibfnamefont {F.}~\bibnamefont {Lorant}}, \ and\ \bibinfo {author} {\bibfnamefont {W.~A.}\ \bibnamefont {Goddard}},\ }\href@noop {} {\bibfield  {journal} {\bibinfo  {journal} {J. Phys. Chem. A}\ }\textbf {\bibinfo {volume} {105}},\ \bibinfo {pages} {9396} (\bibinfo {year} {2001})}\BibitemShut {NoStop}%
\bibitem [{\citenamefont {Liang}\ \emph {et~al.}(2012)\citenamefont {Liang}, \citenamefont {Devine}, \citenamefont {Phillpot},\ and\ \citenamefont {Sinnott}}]{liang2012variable}%
  \BibitemOpen
  \bibfield  {author} {\bibinfo {author} {\bibfnamefont {T.}~\bibnamefont {Liang}}, \bibinfo {author} {\bibfnamefont {B.}~\bibnamefont {Devine}}, \bibinfo {author} {\bibfnamefont {S.~R.}\ \bibnamefont {Phillpot}}, \ and\ \bibinfo {author} {\bibfnamefont {S.~B.}\ \bibnamefont {Sinnott}},\ }\href@noop {} {\bibfield  {journal} {\bibinfo  {journal} {J. Phys. Chem. A}\ }\textbf {\bibinfo {volume} {116}},\ \bibinfo {pages} {7976} (\bibinfo {year} {2012})}\BibitemShut {NoStop}%
\bibitem [{\citenamefont {Hajibabaei}\ \emph {et~al.}(2021{\natexlab{a}})\citenamefont {Hajibabaei}, \citenamefont {Ha}, \citenamefont {Pourasad}, \citenamefont {Kim},\ and\ \citenamefont {Kim}}]{hajibabaei2021machine}%
  \BibitemOpen
  \bibfield  {author} {\bibinfo {author} {\bibfnamefont {A.}~\bibnamefont {Hajibabaei}}, \bibinfo {author} {\bibfnamefont {M.}~\bibnamefont {Ha}}, \bibinfo {author} {\bibfnamefont {S.}~\bibnamefont {Pourasad}}, \bibinfo {author} {\bibfnamefont {J.}~\bibnamefont {Kim}}, \ and\ \bibinfo {author} {\bibfnamefont {K.~S.}\ \bibnamefont {Kim}},\ }\href@noop {} {\bibfield  {journal} {\bibinfo  {journal} {J. Phys. Chem. A}\ }\textbf {\bibinfo {volume} {125}},\ \bibinfo {pages} {9414} (\bibinfo {year} {2021}{\natexlab{a}})}\BibitemShut {NoStop}%
\bibitem [{\citenamefont {Tzanov}\ \emph {et~al.}(2014)\citenamefont {Tzanov}, \citenamefont {Cuendet},\ and\ \citenamefont {Tuckerman}}]{tzanov2014accurately}%
  \BibitemOpen
  \bibfield  {author} {\bibinfo {author} {\bibfnamefont {A.~T.}\ \bibnamefont {Tzanov}}, \bibinfo {author} {\bibfnamefont {M.~A.}\ \bibnamefont {Cuendet}}, \ and\ \bibinfo {author} {\bibfnamefont {M.~E.}\ \bibnamefont {Tuckerman}},\ }\href@noop {} {\bibfield  {journal} {\bibinfo  {journal} {J. Phys. Chem. B}\ }\textbf {\bibinfo {volume} {118}},\ \bibinfo {pages} {6539} (\bibinfo {year} {2014})}\BibitemShut {NoStop}%
\bibitem [{\citenamefont {Kondratyuk}\ \emph {et~al.}(2016)\citenamefont {Kondratyuk}, \citenamefont {Norman},\ and\ \citenamefont {Stegailov}}]{kondratyuk2016self}%
  \BibitemOpen
  \bibfield  {author} {\bibinfo {author} {\bibfnamefont {N.~D.}\ \bibnamefont {Kondratyuk}}, \bibinfo {author} {\bibfnamefont {G.~E.}\ \bibnamefont {Norman}}, \ and\ \bibinfo {author} {\bibfnamefont {V.~V.}\ \bibnamefont {Stegailov}},\ }\href@noop {} {\bibfield  {journal} {\bibinfo  {journal} {J.~Chem.~Phys.}\ }\textbf {\bibinfo {volume} {145}} (\bibinfo {year} {2016})}\BibitemShut {NoStop}%
\bibitem [{\citenamefont {Kondratyuk}(2019)}]{kondratyuk2019comparing}%
  \BibitemOpen
  \bibfield  {author} {\bibinfo {author} {\bibfnamefont {N.}~\bibnamefont {Kondratyuk}},\ }\href@noop {} {\bibfield  {journal} {\bibinfo  {journal} {J.~Phys.:~Conf.~Ser.}\ }\textbf {\bibinfo {volume} {1385}},\ \bibinfo {pages} {012048} (\bibinfo {year} {2019})}\BibitemShut {NoStop}%
\bibitem [{\citenamefont {Ha}\ \emph {et~al.}(2022)\citenamefont {Ha}, \citenamefont {Hajibabaei}, \citenamefont {Pourasad},\ and\ \citenamefont {Kim}}]{ha2022sparse}%
  \BibitemOpen
  \bibfield  {author} {\bibinfo {author} {\bibfnamefont {M.}~\bibnamefont {Ha}}, \bibinfo {author} {\bibfnamefont {A.}~\bibnamefont {Hajibabaei}}, \bibinfo {author} {\bibfnamefont {S.}~\bibnamefont {Pourasad}}, \ and\ \bibinfo {author} {\bibfnamefont {K.~S.}\ \bibnamefont {Kim}},\ }\href@noop {} {\bibfield  {journal} {\bibinfo  {journal} {ACS Phys. Chem. Au}\ }\textbf {\bibinfo {volume} {2}},\ \bibinfo {pages} {260} (\bibinfo {year} {2022})}\BibitemShut {NoStop}%
\bibitem [{\citenamefont {Zhang}\ \emph {et~al.}(2023)\citenamefont {Zhang}, \citenamefont {Mako{\'s}}, \citenamefont {Jadrich}, \citenamefont {Kraka}, \citenamefont {Barros}, \citenamefont {Nebgen}, \citenamefont {Tretiak}, \citenamefont {Isayev}, \citenamefont {Lubbers}, \citenamefont {Messerly},\ and\ \citenamefont {Smith}}]{zhang2023exploring}%
  \BibitemOpen
  \bibfield  {author} {\bibinfo {author} {\bibfnamefont {S.}~\bibnamefont {Zhang}}, \bibinfo {author} {\bibfnamefont {M.}~\bibnamefont {Mako{\'s}}}, \bibinfo {author} {\bibfnamefont {R.}~\bibnamefont {Jadrich}}, \bibinfo {author} {\bibfnamefont {E.}~\bibnamefont {Kraka}}, \bibinfo {author} {\bibfnamefont {K.}~\bibnamefont {Barros}}, \bibinfo {author} {\bibfnamefont {B.}~\bibnamefont {Nebgen}}, \bibinfo {author} {\bibfnamefont {S.}~\bibnamefont {Tretiak}}, \bibinfo {author} {\bibfnamefont {O.}~\bibnamefont {Isayev}}, \bibinfo {author} {\bibfnamefont {N.}~\bibnamefont {Lubbers}}, \bibinfo {author} {\bibfnamefont {R.}~\bibnamefont {Messerly}}, \ and\ \bibinfo {author} {\bibfnamefont {J.}~\bibnamefont {Smith}},\ }\href {\doibase 10.26434/chemrxiv-2022-15ct6-v3} {\enquote {\bibinfo {title} {Exploring the frontiers of chemistry with a general reactive machine learning potential},}\ } (\bibinfo {year} {2023})\BibitemShut {NoStop}%
\bibitem [{\citenamefont {Unke}\ \emph {et~al.}(2021)\citenamefont {Unke}, \citenamefont {Chmiela}, \citenamefont {Sauceda}, \citenamefont {Gastegger}, \citenamefont {Poltavsky}, \citenamefont {Schu{\"{u}}tt}, \citenamefont {Tkatchenko},\ and\ \citenamefont {M{\"{u}}ller}}]{unke2021machine}%
  \BibitemOpen
  \bibfield  {author} {\bibinfo {author} {\bibfnamefont {O.~T.}\ \bibnamefont {Unke}}, \bibinfo {author} {\bibfnamefont {S.}~\bibnamefont {Chmiela}}, \bibinfo {author} {\bibfnamefont {H.~E.}\ \bibnamefont {Sauceda}}, \bibinfo {author} {\bibfnamefont {M.}~\bibnamefont {Gastegger}}, \bibinfo {author} {\bibfnamefont {I.}~\bibnamefont {Poltavsky}}, \bibinfo {author} {\bibfnamefont {K.~T.}\ \bibnamefont {Schu{\"{u}}tt}}, \bibinfo {author} {\bibfnamefont {A.}~\bibnamefont {Tkatchenko}}, \ and\ \bibinfo {author} {\bibfnamefont {K.-R.}\ \bibnamefont {M{\"{u}}ller}},\ }\href@noop {} {\bibfield  {journal} {\bibinfo  {journal} {Chem. Rev.}\ }\textbf {\bibinfo {volume} {121}},\ \bibinfo {pages} {10142} (\bibinfo {year} {2021})}\BibitemShut {NoStop}%
\bibitem [{\citenamefont {Deringer}\ \emph {et~al.}(2021)\citenamefont {Deringer}, \citenamefont {Bart{\'o}k}, \citenamefont {Bernstein}, \citenamefont {Wilkins}, \citenamefont {Ceriotti},\ and\ \citenamefont {Cs{\'a}nyi}}]{deringer2021gaussian}%
  \BibitemOpen
  \bibfield  {author} {\bibinfo {author} {\bibfnamefont {V.~L.}\ \bibnamefont {Deringer}}, \bibinfo {author} {\bibfnamefont {A.~P.}\ \bibnamefont {Bart{\'o}k}}, \bibinfo {author} {\bibfnamefont {N.}~\bibnamefont {Bernstein}}, \bibinfo {author} {\bibfnamefont {D.~M.}\ \bibnamefont {Wilkins}}, \bibinfo {author} {\bibfnamefont {M.}~\bibnamefont {Ceriotti}}, \ and\ \bibinfo {author} {\bibfnamefont {G.}~\bibnamefont {Cs{\'a}nyi}},\ }\href@noop {} {\bibfield  {journal} {\bibinfo  {journal} {Chem. Rev.}\ }\textbf {\bibinfo {volume} {121}},\ \bibinfo {pages} {10073} (\bibinfo {year} {2021})}\BibitemShut {NoStop}%
\bibitem [{\citenamefont {Behler}\ and\ \citenamefont {Parrinello}(2007)}]{behler2007generalized}%
  \BibitemOpen
  \bibfield  {author} {\bibinfo {author} {\bibfnamefont {J.}~\bibnamefont {Behler}}\ and\ \bibinfo {author} {\bibfnamefont {M.}~\bibnamefont {Parrinello}},\ }\href@noop {} {\bibfield  {journal} {\bibinfo  {journal} {Phys. Rev. Lett.}\ }\textbf {\bibinfo {volume} {98}},\ \bibinfo {pages} {146401} (\bibinfo {year} {2007})}\BibitemShut {NoStop}%
\bibitem [{\citenamefont {Pun}\ \emph {et~al.}(2019)\citenamefont {Pun}, \citenamefont {Batra}, \citenamefont {Ramprasad},\ and\ \citenamefont {Mishin}}]{pun2019physically}%
  \BibitemOpen
  \bibfield  {author} {\bibinfo {author} {\bibfnamefont {G.~P.}\ \bibnamefont {Pun}}, \bibinfo {author} {\bibfnamefont {R.}~\bibnamefont {Batra}}, \bibinfo {author} {\bibfnamefont {R.}~\bibnamefont {Ramprasad}}, \ and\ \bibinfo {author} {\bibfnamefont {Y.}~\bibnamefont {Mishin}},\ }\href@noop {} {\bibfield  {journal} {\bibinfo  {journal} {Nat. Commun.}\ }\textbf {\bibinfo {volume} {10}},\ \bibinfo {pages} {2339} (\bibinfo {year} {2019})}\BibitemShut {NoStop}%
\bibitem [{\citenamefont {Batzner}\ \emph {et~al.}(2022)\citenamefont {Batzner}, \citenamefont {Musaelian}, \citenamefont {Sun}, \citenamefont {Geiger}, \citenamefont {Mailoa}, \citenamefont {Kornbluth}, \citenamefont {Molinari}, \citenamefont {Smidt},\ and\ \citenamefont {Kozinsky}}]{batzner20223}%
  \BibitemOpen
  \bibfield  {author} {\bibinfo {author} {\bibfnamefont {S.}~\bibnamefont {Batzner}}, \bibinfo {author} {\bibfnamefont {A.}~\bibnamefont {Musaelian}}, \bibinfo {author} {\bibfnamefont {L.}~\bibnamefont {Sun}}, \bibinfo {author} {\bibfnamefont {M.}~\bibnamefont {Geiger}}, \bibinfo {author} {\bibfnamefont {J.~P.}\ \bibnamefont {Mailoa}}, \bibinfo {author} {\bibfnamefont {M.}~\bibnamefont {Kornbluth}}, \bibinfo {author} {\bibfnamefont {N.}~\bibnamefont {Molinari}}, \bibinfo {author} {\bibfnamefont {T.~E.}\ \bibnamefont {Smidt}}, \ and\ \bibinfo {author} {\bibfnamefont {B.}~\bibnamefont {Kozinsky}},\ }\href@noop {} {\bibfield  {journal} {\bibinfo  {journal} {Nat. Commun.}\ }\textbf {\bibinfo {volume} {13}},\ \bibinfo {pages} {2453} (\bibinfo {year} {2022})}\BibitemShut {NoStop}%
\bibitem [{\citenamefont {Bart{\'o}k}\ \emph {et~al.}(2010)\citenamefont {Bart{\'o}k}, \citenamefont {Payne}, \citenamefont {Kondor},\ and\ \citenamefont {Cs{\'a}nyi}}]{bartok2010gaussian}%
  \BibitemOpen
  \bibfield  {author} {\bibinfo {author} {\bibfnamefont {A.~P.}\ \bibnamefont {Bart{\'o}k}}, \bibinfo {author} {\bibfnamefont {M.~C.}\ \bibnamefont {Payne}}, \bibinfo {author} {\bibfnamefont {R.}~\bibnamefont {Kondor}}, \ and\ \bibinfo {author} {\bibfnamefont {G.}~\bibnamefont {Cs{\'a}nyi}},\ }\href@noop {} {\bibfield  {journal} {\bibinfo  {journal} {Phys. Rev. Lett.}\ }\textbf {\bibinfo {volume} {104}},\ \bibinfo {pages} {136403} (\bibinfo {year} {2010})}\BibitemShut {NoStop}%
\bibitem [{\citenamefont {Chmiela}\ \emph {et~al.}(2017)\citenamefont {Chmiela}, \citenamefont {Tkatchenko}, \citenamefont {Sauceda}, \citenamefont {Poltavsky}, \citenamefont {Sch{\"u}tt},\ and\ \citenamefont {M{\"u}ller}}]{chmiela2017machine}%
  \BibitemOpen
  \bibfield  {author} {\bibinfo {author} {\bibfnamefont {S.}~\bibnamefont {Chmiela}}, \bibinfo {author} {\bibfnamefont {A.}~\bibnamefont {Tkatchenko}}, \bibinfo {author} {\bibfnamefont {H.~E.}\ \bibnamefont {Sauceda}}, \bibinfo {author} {\bibfnamefont {I.}~\bibnamefont {Poltavsky}}, \bibinfo {author} {\bibfnamefont {K.~T.}\ \bibnamefont {Sch{\"u}tt}}, \ and\ \bibinfo {author} {\bibfnamefont {K.-R.}\ \bibnamefont {M{\"u}ller}},\ }\href@noop {} {\bibfield  {journal} {\bibinfo  {journal} {Sci. Adv.}\ }\textbf {\bibinfo {volume} {3}},\ \bibinfo {pages} {e1603015} (\bibinfo {year} {2017})}\BibitemShut {NoStop}%
\bibitem [{\citenamefont {Hajibabaei}\ \emph {et~al.}(2021{\natexlab{b}})\citenamefont {Hajibabaei}, \citenamefont {Myung},\ and\ \citenamefont {Kim}}]{hajibabaei2021sparse}%
  \BibitemOpen
  \bibfield  {author} {\bibinfo {author} {\bibfnamefont {A.}~\bibnamefont {Hajibabaei}}, \bibinfo {author} {\bibfnamefont {C.~W.}\ \bibnamefont {Myung}}, \ and\ \bibinfo {author} {\bibfnamefont {K.~S.}\ \bibnamefont {Kim}},\ }\href@noop {} {\bibfield  {journal} {\bibinfo  {journal} {Phys. Rev. B}\ }\textbf {\bibinfo {volume} {103}},\ \bibinfo {pages} {214102} (\bibinfo {year} {2021}{\natexlab{b}})}\BibitemShut {NoStop}%
\bibitem [{\citenamefont {Hong}\ \emph {et~al.}(2021)\citenamefont {Hong}, \citenamefont {Chun}, \citenamefont {Lee}, \citenamefont {Kim}, \citenamefont {Seo}, \citenamefont {Kang},\ and\ \citenamefont {Han}}]{hong2021first}%
  \BibitemOpen
  \bibfield  {author} {\bibinfo {author} {\bibfnamefont {S.~J.}\ \bibnamefont {Hong}}, \bibinfo {author} {\bibfnamefont {H.}~\bibnamefont {Chun}}, \bibinfo {author} {\bibfnamefont {J.}~\bibnamefont {Lee}}, \bibinfo {author} {\bibfnamefont {B.-H.}\ \bibnamefont {Kim}}, \bibinfo {author} {\bibfnamefont {M.~H.}\ \bibnamefont {Seo}}, \bibinfo {author} {\bibfnamefont {J.}~\bibnamefont {Kang}}, \ and\ \bibinfo {author} {\bibfnamefont {B.}~\bibnamefont {Han}},\ }\href@noop {} {\bibfield  {journal} {\bibinfo  {journal} {J. Phys. Chem. Lett.}\ }\textbf {\bibinfo {volume} {12}},\ \bibinfo {pages} {6000} (\bibinfo {year} {2021})}\BibitemShut {NoStop}%
\bibitem [{\citenamefont {Hajibabaei}\ and\ \citenamefont {Kim}(2021)}]{hajibabaei2021universal}%
  \BibitemOpen
  \bibfield  {author} {\bibinfo {author} {\bibfnamefont {A.}~\bibnamefont {Hajibabaei}}\ and\ \bibinfo {author} {\bibfnamefont {K.~S.}\ \bibnamefont {Kim}},\ }\href@noop {} {\bibfield  {journal} {\bibinfo  {journal} {J. Phys. Chem. Lett.}\ }\textbf {\bibinfo {volume} {12}},\ \bibinfo {pages} {8115} (\bibinfo {year} {2021})}\BibitemShut {NoStop}%
\bibitem [{\citenamefont {Willow}\ and\ \citenamefont {Myung}(2024)}]{willow2024bayesian}%
  \BibitemOpen
  \bibfield  {author} {\bibinfo {author} {\bibfnamefont {S.~Y.}\ \bibnamefont {Willow}}\ and\ \bibinfo {author} {\bibfnamefont {C.~W.}\ \bibnamefont {Myung}},\ }\href@noop {} {\bibfield  {journal} {\bibinfo  {journal} {arXiv preprint arXiv:2402.06256}\ } (\bibinfo {year} {2024})}\BibitemShut {NoStop}%
\bibitem [{\citenamefont {Willow}\ \emph {et~al.}(2024)\citenamefont {Willow}, \citenamefont {Kim}, \citenamefont {Ha}, \citenamefont {Hajibabaei},\ and\ \citenamefont {Myung}}]{willow2024sparse}%
  \BibitemOpen
  \bibfield  {author} {\bibinfo {author} {\bibfnamefont {S.~Y.}\ \bibnamefont {Willow}}, \bibinfo {author} {\bibfnamefont {G.~S.}\ \bibnamefont {Kim}}, \bibinfo {author} {\bibfnamefont {M.}~\bibnamefont {Ha}}, \bibinfo {author} {\bibfnamefont {A.}~\bibnamefont {Hajibabaei}}, \ and\ \bibinfo {author} {\bibfnamefont {C.~W.}\ \bibnamefont {Myung}},\ }\href@noop {} {\enquote {\bibinfo {title} {A sparse bayesian committee machine potential for hydrocarbons},}\ } (\bibinfo {year} {2024}),\ \Eprint {http://arxiv.org/abs/2402.14497} {arXiv:2402.14497 [cond-mat.mtrl-sci]} \BibitemShut {NoStop}%
\bibitem [{\citenamefont {Hinton}(2002)}]{hinton2002training}%
  \BibitemOpen
  \bibfield  {author} {\bibinfo {author} {\bibfnamefont {G.~E.}\ \bibnamefont {Hinton}},\ }\href@noop {} {\bibfield  {journal} {\bibinfo  {journal} {Neural computation}\ }\textbf {\bibinfo {volume} {14}},\ \bibinfo {pages} {1771} (\bibinfo {year} {2002})}\BibitemShut {NoStop}%
\bibitem [{\citenamefont {Tresp}(2000)}]{tresp2000bayesian}%
  \BibitemOpen
  \bibfield  {author} {\bibinfo {author} {\bibfnamefont {V.}~\bibnamefont {Tresp}},\ }\href@noop {} {\bibfield  {journal} {\bibinfo  {journal} {Neural computation}\ }\textbf {\bibinfo {volume} {12}},\ \bibinfo {pages} {2719} (\bibinfo {year} {2000})}\BibitemShut {NoStop}%
\bibitem [{\citenamefont {Bart{\'o}k}\ \emph {et~al.}(2013)\citenamefont {Bart{\'o}k}, \citenamefont {Kondor},\ and\ \citenamefont {Cs{\'a}nyi}}]{bartok2013representing}%
  \BibitemOpen
  \bibfield  {author} {\bibinfo {author} {\bibfnamefont {A.~P.}\ \bibnamefont {Bart{\'o}k}}, \bibinfo {author} {\bibfnamefont {R.}~\bibnamefont {Kondor}}, \ and\ \bibinfo {author} {\bibfnamefont {G.}~\bibnamefont {Cs{\'a}nyi}},\ }\href@noop {} {\bibfield  {journal} {\bibinfo  {journal} {Phys. Rev. B}\ }\textbf {\bibinfo {volume} {87}},\ \bibinfo {pages} {184115} (\bibinfo {year} {2013})}\BibitemShut {NoStop}%
\bibitem [{\citenamefont {Williams}\ \emph {et~al.}(2002)\citenamefont {Williams}, \citenamefont {Rasmussen}, \citenamefont {Scwaighofer},\ and\ \citenamefont {Tresp}}]{williams2002observations}%
  \BibitemOpen
  \bibfield  {author} {\bibinfo {author} {\bibfnamefont {C.}~\bibnamefont {Williams}}, \bibinfo {author} {\bibfnamefont {C.}~\bibnamefont {Rasmussen}}, \bibinfo {author} {\bibfnamefont {A.}~\bibnamefont {Scwaighofer}}, \ and\ \bibinfo {author} {\bibfnamefont {V.}~\bibnamefont {Tresp}},\ }\href@noop {} {\emph {\bibinfo {title} {{Observations on the Nyström Method for Gaussian Process Prediction}}}},\ \bibinfo {type} {Technical Report of the University of Edinburgh}\ (\bibinfo  {institution} {University of Edinburgh},\ \bibinfo {address} {Edinburgh, UK},\ \bibinfo {year} {2002})\BibitemShut {NoStop}%
\bibitem [{\citenamefont {Kresse}\ and\ \citenamefont {Furthm{\"u}ller}(1996)}]{kresse1996efficient}%
  \BibitemOpen
  \bibfield  {author} {\bibinfo {author} {\bibfnamefont {G.}~\bibnamefont {Kresse}}\ and\ \bibinfo {author} {\bibfnamefont {J.}~\bibnamefont {Furthm{\"u}ller}},\ }\href@noop {} {\bibfield  {journal} {\bibinfo  {journal} {Phys. Rev. B}\ }\textbf {\bibinfo {volume} {54}},\ \bibinfo {pages} {11169} (\bibinfo {year} {1996})}\BibitemShut {NoStop}%
\bibitem [{\citenamefont {Bl{\"o}chl}(1994)}]{blochl1994projector}%
  \BibitemOpen
  \bibfield  {author} {\bibinfo {author} {\bibfnamefont {P.~E.}\ \bibnamefont {Bl{\"o}chl}},\ }\href@noop {} {\bibfield  {journal} {\bibinfo  {journal} {Phys. Rev. B}\ }\textbf {\bibinfo {volume} {50}},\ \bibinfo {pages} {17953} (\bibinfo {year} {1994})}\BibitemShut {NoStop}%
\bibitem [{\citenamefont {Perdew}\ \emph {et~al.}(1996)\citenamefont {Perdew}, \citenamefont {Burke},\ and\ \citenamefont {Ernzerhof}}]{perdew1996generalized}%
  \BibitemOpen
  \bibfield  {author} {\bibinfo {author} {\bibfnamefont {J.~P.}\ \bibnamefont {Perdew}}, \bibinfo {author} {\bibfnamefont {K.}~\bibnamefont {Burke}}, \ and\ \bibinfo {author} {\bibfnamefont {M.}~\bibnamefont {Ernzerhof}},\ }\href@noop {} {\bibfield  {journal} {\bibinfo  {journal} {Phys. Rev. Lett.}\ }\textbf {\bibinfo {volume} {77}},\ \bibinfo {pages} {3865} (\bibinfo {year} {1996})}\BibitemShut {NoStop}%
\bibitem [{\citenamefont {Larsen}\ \emph {et~al.}(2017)\citenamefont {Larsen}, \citenamefont {Mortensen}, \citenamefont {Blomqvist}, \citenamefont {Castelli}, \citenamefont {Christensen}, \citenamefont {Du{\l}ak}, \citenamefont {Friis}, \citenamefont {Groves}, \citenamefont {Hammer}, \citenamefont {Hargus} \emph {et~al.}}]{larsen2017atomic}%
  \BibitemOpen
  \bibfield  {author} {\bibinfo {author} {\bibfnamefont {A.~H.}\ \bibnamefont {Larsen}}, \bibinfo {author} {\bibfnamefont {J.~J.}\ \bibnamefont {Mortensen}}, \bibinfo {author} {\bibfnamefont {J.}~\bibnamefont {Blomqvist}}, \bibinfo {author} {\bibfnamefont {I.~E.}\ \bibnamefont {Castelli}}, \bibinfo {author} {\bibfnamefont {R.}~\bibnamefont {Christensen}}, \bibinfo {author} {\bibfnamefont {M.}~\bibnamefont {Du{\l}ak}}, \bibinfo {author} {\bibfnamefont {J.}~\bibnamefont {Friis}}, \bibinfo {author} {\bibfnamefont {M.~N.}\ \bibnamefont {Groves}}, \bibinfo {author} {\bibfnamefont {B.}~\bibnamefont {Hammer}}, \bibinfo {author} {\bibfnamefont {C.}~\bibnamefont {Hargus}},  \emph {et~al.},\ }\href@noop {} {\bibfield  {journal} {\bibinfo  {journal} {J.~Phys.:~Condens.~Matter.}\ }\textbf {\bibinfo {volume} {29}},\ \bibinfo {pages} {273002} (\bibinfo {year} {2017})}\BibitemShut {NoStop}%
\bibitem [{\citenamefont {\textsc{AutoForce}}(2023)}]{AutoForce}%
  \BibitemOpen
  \bibfield  {author} {\bibinfo {author} {\bibnamefont {\textsc{AutoForce}}},\ }\href@noop {} {\enquote {\bibinfo {title} {{A Python Package for Sparse Gaussian Process Regression of the Ab-Initio Potential Energy Surface}},}\ }\bibinfo {howpublished} {https://github.com/amirhajibabaei/AutoForce} (\bibinfo {year} {2023})\BibitemShut {NoStop}%
\end{thebibliography}%
\end{document}


\title{A Bayesian Committee Machine Potential for Organic Nitrogen Compounds}

\author{Hyun Gyu Park}
\affiliation{Department of Energy Science, Sungkyunkwan University, Seobu-ro 2066, Suwon, 16419, Korea}

\author{Soohaeng Yoo Willow}
\affiliation{Department of Energy Science, Sungkyunkwan University, Seobu-ro 2066, Suwon, 16419, Korea}

\author{D. ChangMo Yang}
\email{dcyang@skku.edu}
\affiliation{Department of Energy Science, Sungkyunkwan University, Seobu-ro 2066, Suwon, 16419, Korea}

\author{Chang Woo Myung}
\email{cwmyung@skku.edu}
\affiliation{Department of Energy Science, Sungkyunkwan University, Seobu-ro 2066, Suwon, 16419, Korea}

\date{\today}

\maketitle
\newpage

The following CNH groups were considered for the gas phase:
\begin{itemize}
    \item amine : methylamine, ethylamine, propylamine, butylamine, aniline, phenylamine, isopropylamine, pentanamine, amylamine, hexamine, methenamine, hexanamine, hexylamine, heptanamine, heptylamine octanamine, octylamine, decanamine, decylamine, cyclopropylamine, cyclopentylamine, cyclohexylamine, dimethylamine, ethylmethylamine, diethylamine, naphthylamine, diphenylamine, toluenediamine, trimethylamine, triethylamine, triphenylamine, ethylenediamine, imipramine, desipramine, diethylenetriamine, dimethyltryptamine, tripelennamine, tetraethylenepentamine, 2-aminopyridine, 6-benzylaminopurine, trimipramine, m-phenylenediamine, pheniramine, tryptamine, 3-methyladenine, gramine, 1-methylhistamine, bis(3-aminopropyl)amine, n-methyltryptamine

    \item alkaloid : nicotine, piperidine, sparteine, putrescine, spermidine, spermine, ormosanine, piptanthine, nornicotine, anabasine, anatabine, actinidine, histamine

    \item azole : imidazole, dihydroimidazole, pyrrole, pyrazole, triazole, tetrazole, pentazole, benzimidazole, bifonazole

    \item cyanide : hydrogen-cyanide, n-phenylbenzimidoyl-cyanide, vinylidene-cyanide, cyano-cyanopyridine-cyanide, 2-anilino-1-methylethyl-cyanide, n-ethylpropanimidoyl-cyanide,  2-phenylethanimidoyl-cyanide, hexahydrocarbazole-cyanide

    \item hydrzine : methylhydrazine, mebanazine, ethylhydrazine, dihydralazine, phenylhydrazine, pheniprazine, phenelzine, 1,1-dimethylhydrazine, 1,2-diphenylhydrazine, 1,2-diethylhydrazine, benzylhydrazine, benzophenone-hydrazone, 1-Methyl-1-phenylhydrazine

    \item imidazole: imidazole, benzimidazole, bifonazole, 1-methylimidazole, 1-benzylimidazole, 1-benzyl-2-methylimidazole, 1,2-dimethylimidazole, 2-ethylimidazole, 2-phenylbenzimidazole, 2-ethylbenzimidazole, 2-ethyl-4-methylimidazole, 2-aminobenzimidazole, 2-methylimidazole, 2-methylbenzimidazole, 4-methylimidazole, 5-methylbenzimidazole, 5,6-dimethylbenzimidazole

    \item nitile : adiponitrile, , acrylonitrile, acetonitrile, benzonitrile, butyronitrile, propionitrile, phenylacetonitrile, valeronitrile, heptanenitrile, 3-aminopropionitrile, 3-buteneitrile, 3-phenylpropionitrile, 3-pentenenitrile, 4-pentenenitrile, malononitrile, methacrylonitrile, isobutyronitrile, isovaleronitrile, dodecanenitrile, o-tolunitrile, m-tolunitrile, stearonitrile, tridecanenitrile, trimethylacetonitrile

    \item pyridine : 2-cyanopyridine, 3-cyanopyridine, 4-aminopyridine, 3-cyanopyridine, 3-(pyrrolidin-2-yl)pyridine, phenazopyridine, 2-amino-3-methylpyridine, 2-amino-5-methylpyridine, 2-amino-5-bromopyridine, 2-amino-5-phenylpyridine, 2-amino-6-methylpyridine, 2-amino-4,6-dimethylpyridine, 2-aminopyridine, 3-aminopyridine, 4-aminopyridine, 4-dimethylaminopyridine
\end{itemize}

\newpage 
\begin{figure*} [htp]
\includegraphics[width=6.in]{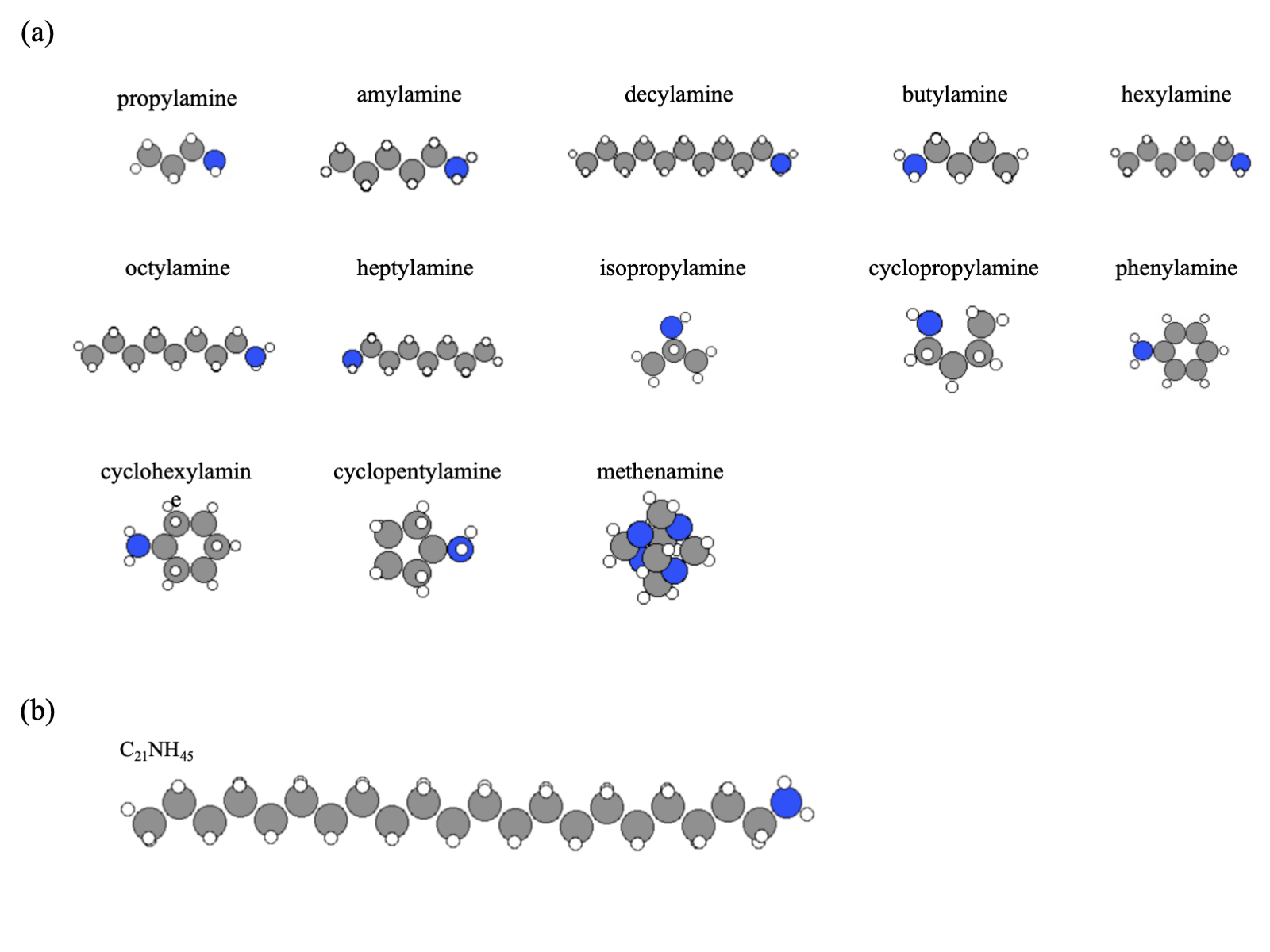}
\caption{(a) illustrates the amine group training dataset utilized to train the ML-IAP model, while 3(b) depicts the amine long chain dataset employed for evaluating the performance of the ML-IAP model.}
\label{fig:schematic}
\end{figure*}

\newpage
\begin{table}[h!]
\caption{Testing the amine long chain for ML-IAP model}

\begin{tabular}{cccc}
\hline
\hline
\ce{C21NH45}                  & \multicolumn{2}{c}{Total   potential energy} &       \\
Time   steps (0.25 fs) & DFT   (eV)          & ML   test (eV)         & MAE   \\
\hline\hline
10000                  & -364.992            & -366.894               & 0.024 \\
20000                  & -365.251            & -365.386               &       \\
30000                  & -365.101            & -365.066               &       \\
40000                  & -364.767            & -364.738               &       \\
50000                  & -364.476            & -364.143               &       \\
60000                  & -364.605            & -364.281               &       \\
70000                  & -364.724            & -364.324               &       \\
80000                  & -364.337            & -364.156               &       \\
90000                  & -364.345            & -363.976               &       \\
100000 (25 ps)         & -364.370            & -363.764               &       \\
\hline   
\end{tabular}
\end{table}